# A CONCURRENCY CONTROL METHOD BASED ON COMMITMENT ORDERING IN MOBILE DATABASES

Ali Karami[1] and Ahmad Baraani-Dastjerdi[2]

[1]Department of Computer Engineering, University of Isfahan, Isfahan, Iran
ali.karami9@yahoo.com
[2] Department of Computer Engineering, University of Isfahan, Isfahan, Iran
ahmadb@eng.ui.ac.ir

## ABSTRACT

*Disconnection of mobile clients from server, in an unclear time and for an unknown duration, due to mobility of mobile clients, is the most important challenges for concurrency control in mobile database with client-server model. Applying pessimistic common classic methods of concurrency control (like 2pl) in mobile database leads to long duration blocking and increasing waiting time of transactions. Because of high rate of aborting transactions, optimistic methods aren`t appropriate in mobile database. In this article, OPCOT concurrency control algorithm is introduced based on optimistic concurrency control method. Reducing communications between mobile client and server, decreasing blocking rate and deadlock of transactions, and increasing concurrency degree are the most important motivation of using optimistic method as the basis method of OPCOT algorithm. To reduce abortion rate of transactions, in execution time of transactions` operators a timestamp is assigned to them. In other to checking commitment ordering property of scheduler, the assigned timestamp is used in server on time of commitment. In this article, serializability of OPCOT algorithm scheduler has been proved by using serializability graph. Results of evaluating simulation show that OPCOT algorithm decreases abortion rate and waiting time of transactions in compare to 2pl and optimistic algorithms.*

## 1. INTRODUCTION

In the recent decade, increase processing ability and decrease in prices of mobile devices have concluded in vast use of these devices in mobile environments. Using data processing services in mobile environments like mobile banking, traffic control and e-commerce, and requirements to easy and swift access to information, are among factors leading to advent of mobile database [1, 2].

Relating to communication process and application of mobile devices, client-server, peer to peer, and adhoc architectures are among most recognized architectures of mobile database. In this study, transactions` concurrency control in mobile database is considered by client-server architecture.

Figure 1 displays client-server architecture. Mobile clients and fix host (server) and mobile base stations are three most important elements of this model. In client-server model, mobile clients are connected to fix host through mobile base station [3].

As has been mentioned in [2] Limited energy sources (battery charge), and unexpected disconnection of mobile client from fix host comprise characteristics of mobile devices. These characteristics aren't a specific fault status, but a general and inherent in mobile environment. However, it has been mentioned in [4] that today, using wireless 802.11 protocol and EVDO method, it is possible to have a synchronized and fast connection between mobile client and server. But, characteristics of mobile devices still impose some limitations upon assessment and data processing in mobile environment. Therefore, for constant and continuous processing, required data are copied partially from server to mobile client.

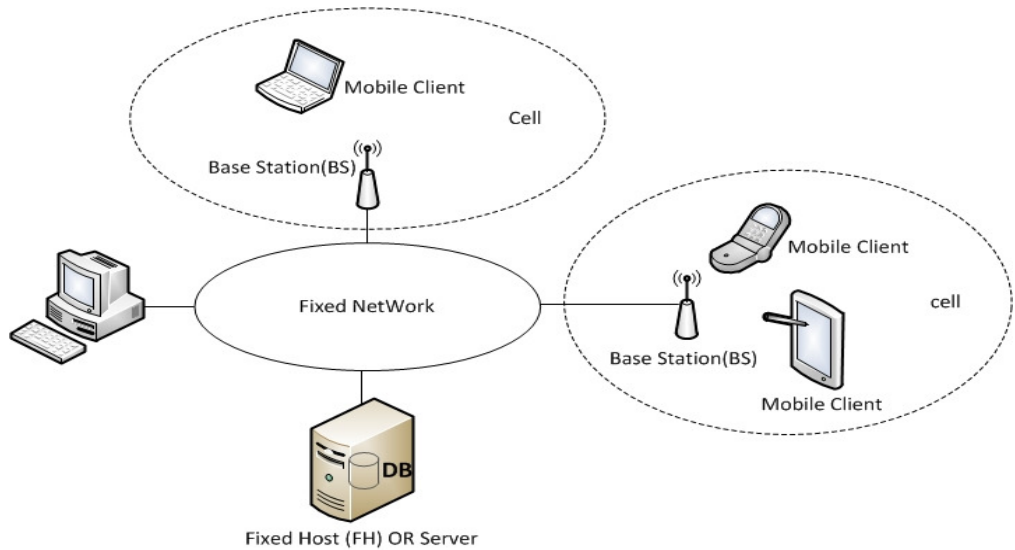

Figure 1. Client-server architecture in mobile database

Concerning aforementioned limitations, concurrency control of concurrent transactions in mobile database face with some difficulties. In pessimistic conventional concurrency protocol based on locking in the time of locking or unlocking, and in protocols based on timestamps, for reading or writing, connection between mobile clients and server is required. In this case non-continuous connection between clients and server led to transactions encounter problems like blocking and long waiting. This leads to inefficiency of pessimistic protocols in mobile environment. In optimistic protocols, no continuous connection between client and server is required. But these protocols are appropriate for optimistic environments in which transaction operators have little conflict. In other non-optimistic environments, using optimistic protocols led to high rate of abortion [2, 3, 5].

In introducing concurrency protocols, issues like reducing rate of abortion, connection between mobile client and server, and time waiting of mobile client, ought to be under considerations. But these measures aren't in accordance in concurrency protocols. In optimistic protocols, communications are less but abortion rate is high; but, in pessimistic ones, the opposite status governs. The main approach in this study is to introduce a concurrency control method based on optimistic concurrency control and commitment ordering schedulers in mobile database. In [6, 7] commitment ordering schedulers for concurrency control has been mentioned.

In the introduced method, just in the beginning and end of transaction, mobile client is connected to server; so, protocols operate optimistically and reduce communications to minimum. Transaction operators and time of operators execution related to time of previous operator execution are added to a list in the transaction execution time. In the time of completing transaction, this list is sent to server; then, based on time of transaction's operation and the concept of commitment ordering schedulers, decisions are made regarding abortion or completion of transaction. This method lessens the amount of aborted transactions significantly. Therefore, in this article, by using timestamp of operators and commitment ordering schedulers, the OPCOT protocol is introduced based on optimistic protocol. This can establish a trade-off between rate of abortion and the amount of connection between mobile client and server. OPCOT protocol doesn't possess the problem of deadlock and blocking of transactions.

This article is organized in five sections. Section two is about previously done researches in the field of concurrency control in mobile database. In section three, we consider the introduced method. The evaluation of protocol is mentioned in section four. Final section talks about conclusion and future works.

## 2. RELATED WORKS

Limitations and specific characteristics of mobile devices cause to little efficiency of conventional concurrency control protocols like 2pl or the other protocols in mobile database. Due to this, various researches have performed in this field. These researches introduce new methods of concurrency control or adapt the existing methods of concurrency control with the requirements of mobile environment.

In section one, it was mentioned that, despite access to fast wireless networks, due to inherent limitations of mobile devices, constant and continuous connection between client and server is not possible through transaction. In locking-based pessimistic protocols, continuous connection of mobile client with server is necessary. If during transaction, this connection will be corrupted, the possibility of unlimited blocking of transactions and also deadlock is appeared. Therefore, based on time out, some researchers have been done regarding methods of non-exclusive locks. In these methods, if transaction doesn't end within the estimated time, locks get caught by transaction will release. References [8, 9] are methods of non-exclusive locks based on time out. In reference [9], a dynamic timer was used to solve the problem of transactions` blocking. In this method, transactions should be finished in specified time, otherwise they would abort. In reference [10] for increasing the efficiency of [9], transactions which don't finish in specified time and are near to final of transaction, wouldn't be aborted. They are allowed to continue execution for specified time. In methods basing on timer (time out) problems like long connection of mobile client with server, estimated the amount of timer, and the length of transaction, do exist. Because of wrong estimation of the required time for completion of transactions, these problems could result in incorrect abortion of transactions. Also, in non-exclusive methods, there is no priority in the order transactions; that is, due to wrong estimation of the required time, a short and less important transaction gets the place of a long and high prioritized one.

A protocol based on AVI is proposed in [11]. This method has still the problem of transaction blocking and computational overhead. This problem has been mentioned in reference [12]. Multi-version concurrency control based on MV2PL protocol and timestamp being introduced in [13] are in fact an extension of method [14].

Hybrid concurrency control protocol of [15] is consisted of optimistic concurrency control protocol and the method based on lock. In this method, it's been attempted to reduce the abortion rate of transactions in optimistic concurrency control by applying weak locking mechanism. But, the number of messages communicates between mobile client and server is high.

In [16, 17], combination of optimistic and pessimistic is performed according to the semantic of operators. In [16], changing two phases locking (2PL) and considering the semantic of transaction`s operators, are executed for increasing the concurrency degree of transactions. If conflicted operators are compatible semantically, then, they can choose a resource simultaneously. In this method, if compatible transactions lock a resource, there would be a resource scarcity for incompatible transaction. Moreover, there is the possibility of high rate of abortion through reconciliation process.

In reference [12] optimistic method is utilized. At the time of completion of transactions, updated data are sent to other mobile transactions using these data. This transfer is in multi-cast status. This technique reduces abortion rate of transactions.

Usually, in wireless networks, bandwidth from mobile client to server is less than the bandwidth from server to mobile client. Reference [18] Puts emphasis on asymmetry of connecting bandwidth between mobile client and server. In [18] optimistic method is under consideration in which in the time of data updating, timestamp of transactions is set dynamically and data is broadcast for mobile clients. Optimistic concurrency control based on timestamp ordering in [19] is adapted for broadcast environments. Also the introduced method in [20] is suitable for broadcast environments.

Increasing concurrency degree, decreasing abortion rate, time waiting, and amount of connection between mobile client and server, lack of deadlock and starvation, reducing computational overhead and communicated data among client and server are among objectives to be considered in concurrency control protocol. Because of the opposite direction of these objectives, in conducted researches, some parameters are optimized. Researches in the field of optimistic concurrency control aim to lessen high rate of transactions` abortion in mobile environment. Pessimistic methods are introduced with the aim of reducing connection between mobile client and server and also transactions blocking in mobile environment.

Considering above purposes, in this study, OPCOT concurrency control method is introduced to increase the concurrency degree of transactions and lessen waiting time of transactions and amount of connection between server and mobile client.

## 3. THE PROPOSED METHOD

The main purpose of this study is to decrease abortion rate of transactions with minimum connection of mobile client and server, and decrease waiting time of mobile clients. To do this, OPCOT algorithm is introduced by assigned timestamp to operators and based on optimistic method according to concept of commitment ordering schedulers.

Transactions are done locally; then, they are sent to server for final stage. In this method, relative timestamps are assigned to operators. Finally, operators` list containing timestamps is transferred to server. The absolute time of operators` execution is calculated by using relative timestamps in server. This is done to ignore effect of local time of mobile clients and synchronize operator timestamps as globally with server time. In server, if the scheduler of transaction is commit order, then transaction would commit successfully.

### 3.1. Commit Ordering Concept

In database concurrency control, commitment ordering is a class of serial scheduler being introduced in reference [6, 7].

If the order of conflict operators is like to completion of transactions, this scheduler is commitment ordering scheduler. For more details, consider transactions $T_J$ and $T_I$. These transactions have conflict, and $T_I$ has priority over $T_J$. If $T_I$ committed before $T_J$, this is a commitment ordering scheduler.

According to [6], every commitment ordering scheduler is a serial scheduler too.

## 3.2. Motivation

In figure 2, it is assumed that transaction $T_J$ aims to commit. It is also assumed this transaction has conflict with $T_I$ and $T_K$ in using common data.

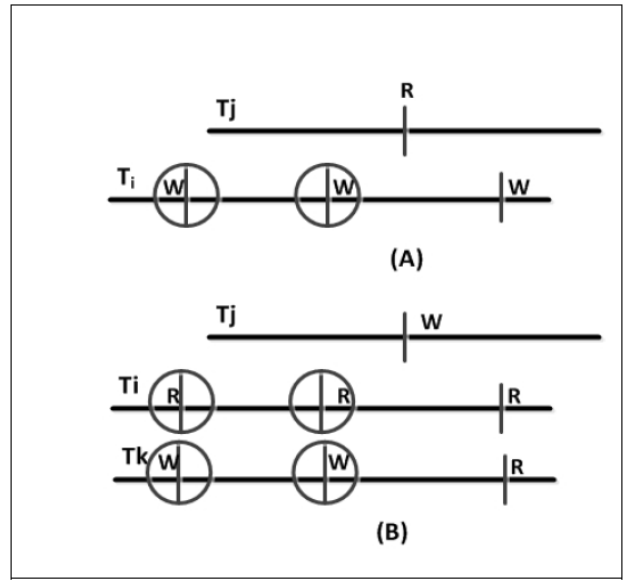

Figure 2. Motivation example

In fig 2-A, transaction $T_J$ and transaction $T_I$ has the conflict of reading with writing. In fig 2-B, transaction $T_J$ has the conflict of writing with reading with transaction $T_I$, and the conflict of writing with writing with transaction $T_K$. In fig 2, in places pointed by circles, $T_J$ are committed by commitment ordering scheduler. But none of these conditions is compatible according to optimistic protocol.

Because of high rate of commitment, we are decided to using commitment order scheduler. Due to no constant and continues connection between mobile client and server throughout transaction execution, timestamps of operators are used to clarify the order of priority of operators. In the time of finalizing transaction, operators` list accompanied by timestamps is utilized to commit or abort transactions according to commitment ordering scheduler in server.

## 3.3. OPCOT algorithm

In this part, the introduced method`s algorithm is presented as pseudo codes. Symbols used in this section and pseudo codes are listed in figure 3.

| $TR_I$ | Transaction I |
|---|---|
| X | X is Data |
| $X . T_R$ | The last time X is read from server |
| $X . T_W$ | The last time X is write on server |
| W(X) | Write X operator |
| R(X) | Read X operator |
| $L_I$ | List of operators and assigned timestamps of transaction I |

Figure 3. The used symbols

In OPCOT protocol, transactions are done in two phases. First phase is reading and execution. In this phase, transactions are able to receive and read all required data from server at the outset of phase. Moreover, to optimize algorithm, transactions can get data through execution time if connection between mobile client and server be available. If data are read within execution time of transaction, the final version of data updated by concurrent transactions would be accessible to transaction. In the case of no connection with server, local data are used. Writings are done locally in first phase, and at the end of first phase, if transaction is performed successfully in the client, the writings for final commitment are sent to server. Because writings are performed at first phase locally without any attention to concurrent transactions, this method is optimistic. High rate of transactions` abortion in non-optimistic environments is one of the most inappropriate characteristics of optimistic methods. The method of this study assigns relative timestamps to every operator during time of execution of operators. These timestamps show the time of each operator relative to time of previous operator. For example, in figure 4, transaction`s operators accompanied by timestamps are listed as $L_I$ = [Begin, 0], [R(X), t1], [W(X), t2], [Commit, t3].

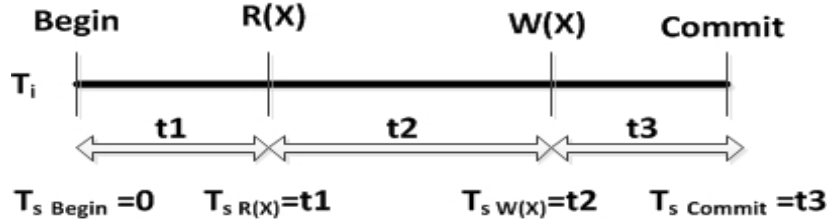

Figure 4. Timestamp example

The list of operators with timestamps in finalization phase is sent to server for commit.

By attention to the possibility of non- synchronization of mobile clients` hours, in introduced algorithm, relative times are used for time stamping of operators. In server, during commitment of transactions, by using relative timestamps, the absolute execution time of operators is calculated proportionate to server time.

Considering these descriptions, pseudo codes of OPCOT concurrency algorithm for mobile client is like figure 5.

1. **Algorithm** Run Transaction On Client

2.     **WHILE** $TR_I$  don't    terminate  **DO**

3.         OP = next operation of $TR_I$

4.         $T_s$ = current time – run time of previous operator

5.         **Add Record** [OP, $T_S$] in $L_I$

6.         **RUN** (OP)

7.     **ENDWHILE**

8.     **SEND** $L_I$ to server for commit $TR_I$

Figure 5. The algorithm of performing transaction in mobile client

After executing each operator, the relative time of this operator is calculated in line 4 in variable Ts. Timestamp calculation is done with regard to previous operator`s time. Operators assign timestamps in line 5. At the end of transaction, in line 8, operators list with timestamps is sent to server for commitment.

After finalize transaction $T_I$ in mobile client, $L_I$ list is sent to server. In server, with regard to time of receiving $L_I$ list and relative timestamps of this list, time of execution operators is calculated according to algorithm presented in figure 6. This process is required because of lack of synchronization in clock of mobile clients and server. In the loop of line 6, absolute timestamp is calculated based on relative timestamps. In this algorithm, absolute timestamps are held in $L_{SI}$ list.

1. **Algorithm** Setting Timestamps Based On Server Time

2.     N: number of transaction operations

3.     $L_{SI}$ [N]. $T_S$ =current time of server

4.     **WHILE**  N>0 **DO**

5.         N=N-1

6.         $L_{SI}$ [N]. $T_S$ = $L_{SI}$ [N+1]. $T_S$ – $L_I$ [N+1]. $T_S$

7.     **ENDWHILE**

8.     $L_I$= $L_{SI}$

Figure 6: The algorithm of setting timestamps based on server time

For every data of X, two timestamps are used. $T_R$ shows the last time X is read from server and $T_W$ the last time when X is written on server
.
When the first phase of transaction in mobile client is completed successfully, $L_I$ list is sent to server for commitment. In algorithm of figure 7, server with regard to $L_I$ list related to $TR_I$ transaction, decides to either commit or abort transaction. If scheduler including $TR_I$ and committed transactions be a commitment ordering scheduler, $TR_I$ transaction will commit; otherwise, it will abort. Commitment ordering is considered by $T_R$ and $T_W$ stamps of X`s data.

In lines 6 and 7, this issue is shown. If at the time of Ts, transaction reads X data locally, and if $T_W$ less than Ts, because the order of conflicted operators is same to the order of commitment, in this case transactions in scheduler keep the characteristics of commitment ordering. Lines 13 and 14 are about considering commitment ordering of writing. In the case of keep the characteristics of commitment ordering scheduler, timestamps of X data are updated, and transaction will complete. Otherwise, there would be an abortion in transaction, and results roll back.

```
1. Algorithm Commit Transaction On Server
2.     CALL   Setting Timestamps Based On Server Time
3.     flag= TRUE
4.     WHILE   L_I is not empty AND flag DO
5.         Rec = get next record of L_I
6.         IF Rec . OP = = R (X) THEN
7.             IF Rec . T_s < X . T_w THEN
8.                 flag=FALSE
9.             ELSE
10.                 X.T_R= Rec . T_s
11.             ENDIF
12.         ENDIF
13.         IF Rec . OP = = W (X) THEN
14.             IF Rec . T_s < X . T_w OR  Rec . T_s < X . T_R THEN
15.                 flag=FALSE
16.             ELSE
17.                 X.T_w= Rec . T_s
18.             ENDIF
19.         ENDIF
20.     ENDWHILE
21.     IF flag THEN
22.         transaction is committed
23.     ELSE
24.         rollback transaction effects
25.         transaction is aborted
26.     ENDIF
```

Figure 7. The algorithm of commit transaction on server

### 3.4. Proving the serializability of OPCOT algorithm

In introduced algorithm, the list of transactions` operators having timestamps is sent to server for commitment. We will approve that only if scheduler involves sent transaction and completed transactions are preserved serializability characteristics, then the transaction of mobile client would be committed; otherwise, it would be aborted.

Suppose $TR_I$ transaction is the last one which successfully performs reading over X at $T_{RI(X)}$ time and writing at $T_{WI(X)}$ time. This way, timestamp for X data would be TR= $T_{RI(X)}$ and TW= $T_{WI(X)}$. For each transaction $T_J$ which is sent to server after $T_I$ and gets access to X data, the following conditions are possible:

The first situation:

If $T_J$ reads from X data at the time of $T_{RJ(X)}$, here the operator should be considered by read to write conflict. This situation is studied according the algorithm presented in figure 7. If there is X.TW <= $T_{RJ(X)}$, the reading will be accepted. Therefore:

X.TW <= $T_{RJ(X)} \equiv T_{WI(X)}$ <= $T_{RJ(X)}$

That is, in serializability graph, there is an edge like $T_I \rightarrow T_J$. (1)

The second situation:

If $T_J$ writes to X data at the time of $T_{WJ(X)}$, then this operator ought to be considered through writing to reading conflict and writing to writing conflict. According to figure 7`s algorithm, only if line 14 be in condition of X.TR <=$T_{WJ(X)}$ and X.TW <=$T_{WJ(X)}$ , then this kind of writing is acceptable. So,

X.TW <= $T_{WJ(X)} \equiv T_{WI(X)}$ <= $T_{WJ(X)}$

X.TR <= $T_{WJ(X)} \equiv T_{RI(X)}$ <= $T_{WJ(X)}$

In this case, in serializability graph, there can be an edge like $T_I \rightarrow T_J$ (2)

Results of (1) and (2) clarify that in all situations in which between TI and TJ in serializability graph is an edge, the direction of edge would be $T_I \rightarrow T_J$. we supposed that $T_J$ is committed after $T_I$, so the direction of edge in the introduced algorithm`s serializability graph represents the commitment ordering of transactions.

That is: $T_I \rightarrow T_J$    $\rightarrow$    Commit $(T_I)$ < Commit $(T_J)$ (3)

Theorem: serializability characteristic of scheduler is preserved by OPCOT algorithm.

To prove this theorem, we show that serializability graph based on OPCOT algorithm doesn't involve loop. This issue will be approved by contradiction proof.

Contradiction proof: assume that in serializability graph, there is a loop like the following:

$T_I \rightarrow ... \rightarrow T_J \rightarrow T_K \rightarrow T_I$

In this way, according to result (3):

Commit $(T_I)$ <Commit $(T_K)$ <Commit $(T_J)$ < ...< Commit $(T_I)$

Therefore, because Commit $(T_I)$ can`t be smaller than Commit $(T_I)$, this condition is impossible. So the assumption is wrong and the theorem is proved.

## 4. EVALUATION OF THE INTRODUCED METHOD

In this section, the method of concurrency control discussed in this study would be evaluated. To do this, it is compared with concurrency control methods of optimistic and 2PL. Because OPCOT protocol is based on optimistic methods hence optimistic concurrency control protocol is selected for comparison and evaluation of introduced protocol. Indeed OPCOT protocol is based on optimistic methods that use commitment ordering scheduler with timestamp technique for transaction operators. Also the proposed protocol is compared with pessimistic protocols (2PL).

In order to evaluate and compare performance of OPCOT, optimistic and 2PL these protocols will be simulated. According to our knowledge, there aren't any simulators for concurrency control simulation in mobile databases. Due to effect of mobile clients movement on concurrency control, selected simulator must support mobile networks and also be able to implement user defined applications and protocols. We use open source software of network simulator NS-3 [21, 22]. This software has the ability of defining new protocol in various network layers, and also supports implementation of mobile networks.

For simulation, we use NS-3 version 3.11 that was installed on Linux Ubuntu 11.04 operating system in dell laptop with Intel(R) Core(TM) 2 Duo CPU 2.00GHz and 4 GB of memory.

Simulation Can be done using Python or C + + languages in the NS3 simulator. In order to add user defined protocols in NS3 layers, protocols were designed as classes. These classes must be inherited from related classes in corresponding layers and related function must be overwritten [21, 22]. Hence, the considered protocols, after implementing by C++ language, were added to NS3 as application layer protocols. For example we implemented a class for OPCOT protocol in NS3 application layer. This class inherited from NS3 application class.

After adding intended protocols to NS3 simulator, mobile networks simulation with the following characteristics was done. A square-shaped area with 2000×2000 meters space, a fix host (server), 10 mobile base stations, and 500 mobile clients are the characteristics of simulated mobile network. Also implemented protocols that were added to NS3 are used by clients and server. Clients move randomly in different directions and communicate their requirements by mobile base station to the server.

Read and write instructions (with R(X) and W(X) notation in this paper) are Instructions which were used in the transaction. The number of instructions in each transaction, which is performed by mobile clients, is considered as floor of normal random variable with mean of 50 instructions and variance of 10 N (50, 10). The amount of reading and writing orders is instructions equal in each transaction.

In server, a table with 1000 data entity was considered. Also, to evaluate the effect of conflict rate in evaluation parameters, in addition to simulation with 1000 data entity, above mentioned mobile network was simulated with 10000 data entity in server. With the same transaction when the number of data entity in server was increased, the conflict rate was decreased, because probability of using same data with concurrent transactions will be reduced.

For each evaluated method, 100 runs with various numbers of transactions were done. Results of simulation are presented in the following pages as plot. For better evaluation, by using regression model, an approximate function is fitted to points. The $R^2$ coefficient of each fitted model is greater than %80. Plots and regression models were created by R. R is open source statistical analysis software. reference [23] is a quick guide about R software.

### 4.1. Abortion rate of transactions

One important evaluated parameter is the rate of abortion of transactions. In figure 8, for 1000 entity, the number of abortions is plotted in front of the number of transactions. Also in figure 9, for 10000 entities, this parameter is evaluated.

Each point on the plot indicates a run. For example the point at coordinate (200, 10) which is related to OPCOP protocol, indicates that in run with 200 transactions, 10 transactions were aborted.

By increasing the number of transactions, due to the possibility of access to a similar data, the number of abortion increases. As it is clear from figure 8 and figure 9, the number of abortion in optimistic algorithm is more than 2PL algorithm. The number of abortion in these two algorithms is more than OPCOT algorithm.

By comparing figure 8 and 9, it is concluded that decreasing conflict rate of transactions operators lead to number of transaction abortion reduction. This conclusion is reasonable.

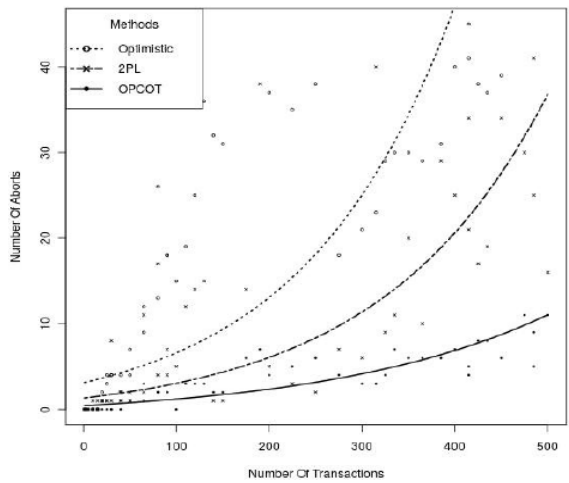

Figure 8. Number of abortion plot (number of entities entity are 1000)

Also abortion rate is calculated as following.

Abortion rate = (Number of transactions abort in all runs) / (Number of transactions in all runs)

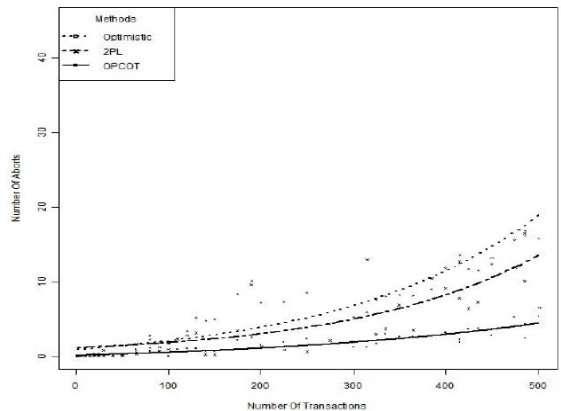

Figure 9. Number of abortion plot (number of data entities are 10000)

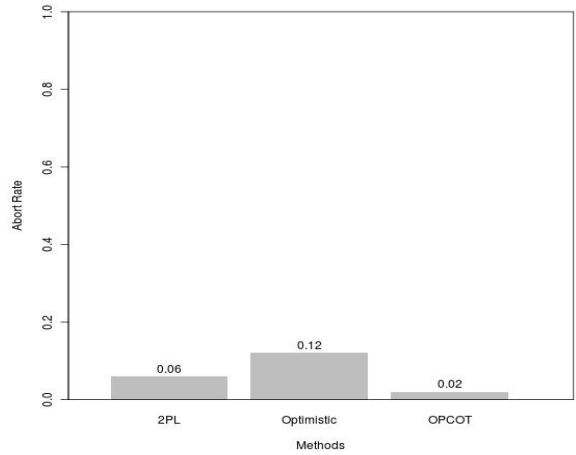

Figure 10. Bar plot of abortion rate (number of data entities are 1000)

For better comparisons, in figure 10 and 11, bar plot of abortion rates for 1000 and 10000 data entities are presented.

Results of figure 8 and 9 are totally approvable in figure 10 and 11.

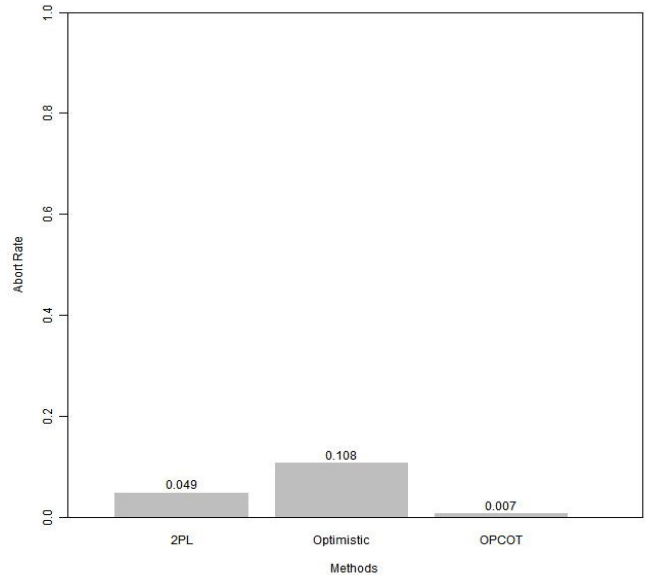

Figure 11. Bar plot of abortion rate (number of data entities are 10000)

## 4.2. Waiting time of transactions

One more significant factor being evaluated is waiting time of each transaction. It is calculated as follows:

Waiting time = time of transaction execution- length of transaction

Length of transaction is the sum of time of performing the instruction of transaction.

Results of this evaluation are presented in figure 12 and 13 for 1000 and 10000 data entities respectively. As you can see, 2Pl algorithm has the longest waiting time. It is acceptable because in this algorithm, for each instruction, there should be a connection to server for locking. Optimistic and OPCOT algorithms, at the beginning, have small waiting time; but, by increasing the number of transactions, the distance between waiting time of these two algorithms exceeds.

Therefore, the growth of waiting time in OPCOT algorithm will be less than optimistic one. Steep growth of plots in figure 13 is less than figure 12. This means that decreasing conflict rate leads to decreasing wait time of transactions. This conclusion is true and acceptable.

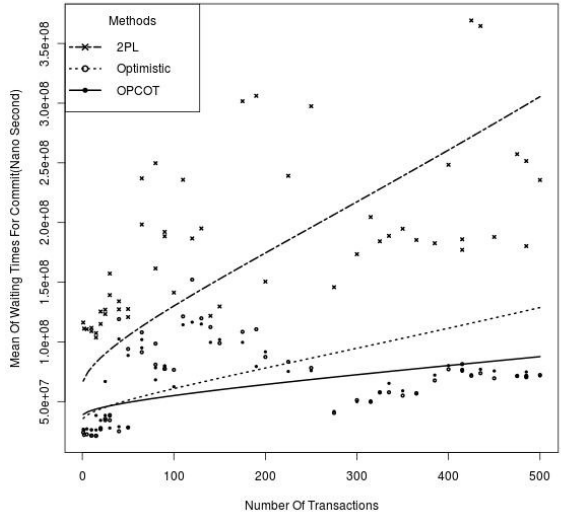

Figure 12: waiting time plot (number of data entities are 1000)

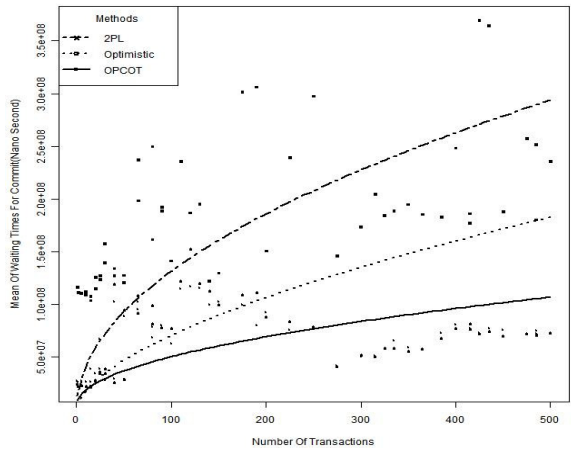

Figure 13: waiting time plot (number of data entities are 10000)

OPCOT algorithm doesn't have deadlock like 2PL. In OPCOT, the amount of connection with the server is less and just at the beginning and end of transaction.

The main problem of OPCOT is that in this algorithm, for each instruction of reading and writing, a timestamp is assigned to instruction. These data should be sent to server in completion of transaction. So, the amount of transferred data between server and mobile client imposes an overhead to system.

## 5. CONCLUSION AND FUTURE WORKS

In this article, OPCOT concurrency control based on optimistic method and commitment ordering schedulers in mobile database with client-server model is introduced. This method is based on optimistic method, so there is no blocking or deadlock in transactions. Moreover, concurrency degree of transactions is high. To reduce the rate of abortion, OPCOT algorithm, based on commitment ordering schedulers, applies the technique of operators` time stamping. This leads to increase in the rate of commitment of transactions. Serializability of scheduler in introduced method was approved. The main problem of OPCOT method is overhead of timestamps and their calculations. This overhead is worthless and is tolerable. We plan to improve this method by introducing it in multi-version status.

## Authors

**Ali Karami** is M.Sc. student of software engineering at the Faculty of Engineering of the University of Isfahan (UI). He earned his B.Sc. degrees from the University of Arak His research interests are mobile databases and distributed systems.

**Ahmad Baraani-Dastjerdi** is an assistant professor of computer engineering at the School of Engineering of the University Of Isfahan (UI). He got his BS in Statistics and Computing in 1977. He got his M.Sc. & PhD degrees in Computer Science from George Washington University in 1979 &University of Wollongong in 1996, respectively. He is Head of the Research Department of the Communication systems and Information Security (CSIS) and Head of the ACM International Collegiate Programming Contest (ACM/ICPC) of University of Isfahan from 2000 until present. He co-authored three books in Persian and received an award of "the Best e-Commerce Iranian Journal Paper" (2005). Currently, he is teaching PhD and MS courses of Advance Topics in Database, Data Protection, Advance Databases, and Machining Learning. His research interests lie in Databases, Data security, Information Systems, e-Society, e-Learning, e-Commerce, Security in e-Commerce, and Security in e-Learning.